\documentclass{elsart}

\usepackage{cite}

\usepackage{dcolumn}

\usepackage{graphicx}

\begin{document}

\begin{frontmatter}

\title{Rational approximation to Thomas--Fermi equations}

\author{Francisco M. Fern\'{a}ndez \thanksref{FMF}}

\address{INIFTA (UNLP,CCT La Plata-CONICET), Divisi\'{o}n Qu\'{i}mica Te\'{o}rica,\\
Diag. 113 y 64 (S/N), Sucursal 4, Casilla de Correo 16,\\
1900 La Plata, Argentina}

\thanks[FMF]{e--mail: fernande@quimica.unlp.edu.ar}

\begin{abstract}
We show that a simple and straightforward rational approximation to the
Thomas--Fermi equation provides the slope at origin with unprecedented
accuracy and that relatively small Pad\'e approximants are far more accurate
than more elaborate approaches proposed recently by other authors.
We consider both the Thomas--Fermi equation for isolated
atoms and for atoms in strong magnetic fields.
\end{abstract}

\end{frontmatter}

\section{Introduction\label{sec:Intro}}

The Thomas--Fermi (TF) equation has proved useful for the treatment of many
physical phenomena that include atoms\cite{BCR74,CM50,M57,MT79,M83},
molecules\cite{M52,M57}, atoms in strong magnetic
fields\cite{BCR74,MT79,M83}, crystals\cite{UT55} and
dense plasmas\cite{YK89} among others. It is well--known that an accurate
solution to that equation is based on the accurate calculation
of the slope at
origin\cite{KMNU55,PP87,FO90}. In particular we mention the rational
approximation in terms of Pad\'e
approximants\cite{T91,EFGP99} because it is relevant to present
discussion. As expected from the great physical significance of the TF equation,
its mathematical aspects have been studied in detail\cite{H69,H70,LS77,L81}.

In spite of being a quite old problem in theoretical physics,
there has recently been a renewed interest in analytical solutions to the
TF equation. For example, Liao\cite{L03} and later Khan and Xu\cite{KX07}
proposed the application of the so called homotopy analysis method (HAM).
More recently Parand and Shahini\cite{PS09} showed that a pseudospectral method
based on Chebyshev polynomials is more accurate than HAM.
Unfortunately, Parand and Shahini\cite{PS09} were not aware
that we had earlier shown that the well--known
straightforward Pad\'e approximants are much more accurate than HAM\cite{F08}.
Therefore, they compared their pseudospectral results with the rather
insufficiently accurate HAM ones that do not constitute a suitable benchmark.

The purpose of this paper is twofold: first we compare the results of the simple
and straightforward Hankel--Pad\'e method (HPM)\cite{F08} with the supposedly accurate
Chebyshev pseudospectral ones\cite{PS09}. Second, we show that the HPM
also gives accurate results for the TF equation for atoms in strong magnetic fields
that has not been treated before in this way.

In Section~\ref{sec:TFeq} we briefly introduce the TF equations for both problems and show
how to transform them into more tractable differential equations. In Section~\ref{sec:HPM}
we outline the main ideas behind the HPM, apply it to both TF equations, and compare
HPM and Chebyshev pseudospectral results for isolated atoms. In
Section~\ref{sec:conclusions} we summarize the main results and comment on other approaches
to nonlinear ordinary differential equations.

\section{The Thomas--Fermi equation \label{sec:TFeq}}

The dimensionless form of the TF equation for atoms\cite{BCR74,CM50,M57,MT79,M83}
\begin{equation}
u^{\prime \prime }(x)=\sqrt{\frac{u(x)^{3}}{x}},\;u(0)=1,\;u(\infty )=0
\label{eq:TF}
\end{equation}
is an example of two--point nonlinear boundary--value problem. When solving
this ordinary differential equation one faces the calculation of the slope
at origin $u^{\prime }(0)$ that is consistent with the physical boundary
conditions indicated in equation (\ref{eq:TF}).

In order to make this letter self contained we review the main features of the
HPM\cite{F08}. It is convenient to define the
variables $t=x^{2}$ and $f(t)=u(t^{2})^{1/2}$, so that the TF equation
becomes
\begin{equation}
t\left[ f(t)f^{\prime \prime }(t)+f^{\prime }(t)^{2}\right] -f(t)f^{\prime
}(t)-2t^{2}f(t)^{3}=0  \label{eq:TFb}
\end{equation}
We expand the solution $f(t)$  in a Taylor series about $t=0$:
\begin{equation}
f(t)=\sum_{j=0}^{\infty }f_{j}t^{j}  \label{eq:f_series}
\end{equation}
where the coefficients $f_{j}$ depend on the only unknown one
$f_{2}=f^{\prime \prime}(0)/2=u^{\prime }(0)/2$.
On substituting the
series (\ref{eq:f_series}) into equation (\ref{eq:TFb}) we easily calculate
as many coefficients $f_{j}$ as desired; for example, the first ones are
\begin{equation}
f_{0}=1,\;f_{1}=0,\;f_{3}=\frac{2}{3},\;f_{4}=-\frac{f_{2}^{2}}{2},\;f_{5}=-%
\frac{4f_{2}}{15},\ldots
\end{equation}

The TF equation for atoms in strong magnetic fields is somewhat simpler\cite
{BCR74,MT79,M83}
\begin{equation}
u^{\prime \prime }(x)=\sqrt{xu(x)},\;u(0)=1,\;u(\infty )=0  \label{eq:TF2}
\end{equation}
By means of the same change of variables we obtain
\begin{equation}
t\left[ f(t)f^{\prime \prime }(t)+f^{\prime }(t)^{2}\right] -f(t)f^{\prime
}(t)-2t^{4}f(t)=0  \label{eq:TF2b}
\end{equation}
as well as the solution in the form of the Taylor series (\ref{eq:f_series})
with coefficients
\begin{equation}
f_{0}=1,\;f_{1}=0,\;f_{3}=0,\;f_{4}=-\frac{f_{2}^{2}}{2},\;f_{5}=\frac{2}{15}%
,\ldots
\end{equation}
where $f_{2}=u^{\prime }(0)/2$ as in the preceding case.

\section{The Hankel--Pad\'e method \label{sec:HPM}}

The HPM is based on the transformation of the power series (\ref{eq:f_series}%
) into a rational function or Pad\'{e} approximant
\begin{equation}
\lbrack M/N](t)=\frac{\sum_{j=0}^{M}a_{j}t^{j}}{\sum_{j=0}^{N}b_{j}t^{j}}
\label{eq:[M/N]}
\end{equation}
One would expect that $M<N$ in order to have the correct limit at infinity;
however, in order to obtain an accurate value of $f_{2}$ it is more
convenient to choose $M=N+d$, $d=0,1,\ldots $ as in previous applications of
the approach to the Schr\"{o}dinger equation (in this case it was called
Riccati--Pad\'{e} method (RPM))\cite
{FMT89,F92,FG93,F95,F95b,F95c,F96,F96b,F97}.

The rational function (\ref{eq:[M/N]}) has $2N+d+1$ coefficients that we may
choose so that $T([M/N],t)=\mathcal{O}(t^{2N+d+1})$, where $T(f,t)=0$ stands
for either equation (\ref{eq:TFb}) or (\ref{eq:TF2b}), and in both cases the
coefficient $f_{2}$ remains undetermined. If we require that $T([M/N],t)=%
\mathcal{O}(t^{2N+d+2})$ we have another equation from which we obtain $f_{2}
$. However, it is simpler and more practical to proceed in a different (and
entirely equivalent) way and require that
\begin{equation}
\lbrack M/N](t)-\sum_{j=0}^{2N+d+1}f_{j}t^{j}=\mathcal{O}(t^{2N+d+2})
\label{eq:[M/N]2}
\end{equation}
In order to satisfy this condition it is necessary that the Hankel
determinant vanishes
\begin{equation}
H_{D}^{d}=\left| f_{i+j+d+1}\right| _{i,j=0,1,\ldots N}=0,  \label{eq:Hankel}
\end{equation}
where $D=N+1$ is the dimension of the Hankel matrix. Each Hankel determinant
is a polynomial function of $f_{2}$ and we expect that there is a sequence
of roots $f_{2}^{[D,d]}$, $D=2,3,\ldots $ that converges towards the actual
value of $u^{\prime }(0)/2$ for a given value of $d$. We compare sequences
with different values of $d$ for inner consistency (all of them should give
the same limit).

Present approach is simple and straightforward: we just obtain the Taylor
coefficients $f_{j}$ from the differential equations (\ref{eq:TFb}) or (\ref
{eq:TF2b}) in terms of $f_{2}$, construct the Hankel determinant, and calculate
its roots. Since $f_{4}$ is the first nonzero coefficient that depends on $%
f_{2}$ we choose Hankel sequences with $d\geq 3$.

The Hankel determinant $H_{D}^{d}$ exhibits many roots and their number
increases with $D$. If we compare the roots of $H_{D}^{d}$ with those of $%
H_{D-1}^{d}$ we easily identify the sequence $f_{2}^{[D,d]}$ that converges
towards $u^{\prime }(0)/2$.

Present HPM may be considered to be a systematic generalization of Tu's
approach\cite{T91} and is clearly different from the strategy proposed by
Epele et al\cite{EFGP99}. We stress that we have
been using the Hankel condicion~(\ref{eq:Hankel}) for quite a long
time\cite {FMT89,F92,FG93,F95,F95b,F95c,F96,F96b,F97}.

We first consider the TF equation for atoms. The convergence of the HPM
has already been discussed in our earlier paper\cite{F08}; therefore, here we
simply compare our results with those of Parand and Shahini\cite{PS09}. Our
estimate of $u^{\prime }(0)$ $-1.588071022611375313$ is by far more accurate
than the one obtained by Parand and Shahini\cite{PS09} $-1.5880702966$,
Liao\cite{L03} $-1.58606$ and Khan and Xu\cite{KX07} $-1.586494973$. Notice
that all those authors kept several inaccurate  digits in their results. This
practice is misleading because it suggests that the calculations are more
accurate than what they really are.

Table~\ref{tab:u(x)} shows our earlier results\cite{F08}, those of Parand and
Shahini\cite{PS09} and the numerical calculation of
Kobayashi et al\cite{KMNU55}. Those HPM results obtained from a relatively small
$[5/8]$ Pad\'e approximant are more accurate than the numerical ones
for $x\leq 1$. We appreciate that the simple and
straightforward Pad\'e approximants are far more accurate than the more
elaborated pseudospectral approach\cite{PS09}. Notice that also in this case
Parand and Shahini\cite{PS09} kept several misleading inaccurate digits.

\begin{figure}[H]
\begin{center}
\includegraphics[width=9cm]{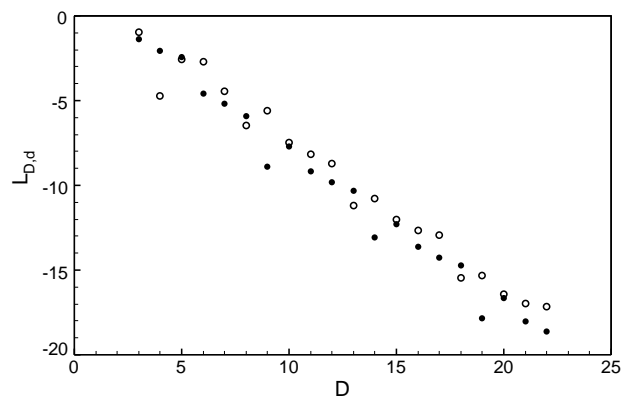}
\end{center}
\caption{$L_{D,d}=\log \left| 2f_{2}^{[D,d]}-2f_{2}^{[D-1,d]}\right| $ for $%
d=4$ (circles) and $d=5$ (filled circles) }
\label{fig:logconv2}
\end{figure}

Fig. \ref{fig:logconv2} shows
$L_{D,d}=\log \left| 2f_{2}^{[D,d]}-2f_{2}^{[D-1,d]}\right| $ for
$D=3,4,\ldots $ that provides a reasonable indication of the convergence of
the sequence of roots for the TF equations for atoms
in strong magnetic fields. We clearly appreciate the great convergence rate of
the sequences with $d=4$ and $d=5$ (the rate of convergence for the case $d=3$
is slightly smaller but still suitable for practical applications).
From those sequences we
estimate $u^{\prime }(0)=-0.93896688764395889306$ that is exact to the last
digit and considerably more accurate than the results published earlier\cite
{BCR74,MT79,M83}.

\section{Conclusions \label{sec:conclusions}}
In this letter we have shown that the well--known simple and straightforward
Pad\'e approximants provide much more accurate results for the TF equation
than the more elaborated Chebyshev pseudospectral method\cite{PS09}, exactly
as we did earlier with the rather cumbersome HAM\cite{F08}. We also applied
the HPM to the TF equation for an atom in a strong magnetic field, analyzed
the convergence of the Hankel sequences towards the slope at origin, and
obtained its value with unprecedented accuracy.

The TF equation is an example of two--point boundary value problems that are
most important in theoretical physics. We have already applied the HPM to other
such problems\cite{AF07}. There are alternative approaches for the accurate
treatment of two--point boundary value problems.
Here we mention the work of Boisseau
et al\cite{BFG07}, Bervillier et al\cite{BBG08,BBG08b}, and in particular
a recent comprehensive discussion of power--series methods for ordinary
differential equations\cite{B08}.

\begin{table}[H]
\caption{Values of the Thomas--Fermi function $u(x)$ obtained by present
method, the Chebyshev pseudospectral one\cite{PS09},
and numerical integration\cite{KMNU55}}
\label{tab:u(x)}
\begin{center}
\par
\begin{tabular}{D{.}{.}{2}D{.}{.}{10}D{.}{.}{10}D{.}{.}{10}}
\hline
\multicolumn{1}{c}{$x$}& \multicolumn{1}{c}{HPM}&
\multicolumn{1}{c}{Chebyshev}& \multicolumn{1}{c}{Numerical} \\
\hline
 1  & 0.424008   &   0.424333179  &  0.42401      \\
 5  & 0.078808   &   0.078277758  &  0.078808     \\
10  & 0.024315   &   0.025044744  &  0.024314     \\
20  & 0.005786   &   0.006585633  &  0.0057849      \\
50  & 0.000633   &   0.000761317  &  0.00063226        \\
100 &0.0001005   &   0.000023409  &  0.00010024       \\

\hline

\end{tabular}
\par
\end{center}
\end{table}

\end{document}